\documentclass[cameraready]{Interspeech}

\title{CrossAccent-TTS: Cross-Lingual Accent-Intensity Controllable Text-to-Speech via Disentangled Speaker and Accent Representations
}

\author[ orcid=0009-0000-5817-9750, equalcontribution]{Ram}{Annamdevula}
\author[ orcid=0009-0000-8450-7797, equalcontribution]{Ankit}{Tatawat}
\author[ orcid=0000-0003-3590-503X]{Ashishkumar P.}{Gudmalwar} 
\author[ orcid=0000-0002-7294-6757]{\\[3pt]Nirmesh J.}{Shah} 
\author[ orcid=0000-0001-5602-2901]{Pankaj}{Wasnik} 

\address{
    Media Analysis, Sony Research India}

\email{\{ram.annamdevula,ankit.tatawat,ashish.gudmalwar1,nirmesh.shah,pankaj.wasnik\}@sony.com}

\keywords{LLM based TTS, Controllable Speech Synthesis, Accent, Accent Conversion}

\usepackage{comment}
\usepackage{xurl}

\begin{document}

\maketitle

\begin{abstract}
    Accent conversion and controllability remain fundamental challenges in cross-lingual text-to-speech (TTS), particularly for low-resource and phonetically diverse Indic languages. While recent large language model (LLM)-based TTS systems exhibit strong cross-lingual generalization, they provide limited explicit control over accent characteristics and intensity. In this paper, we propose CrossAccentTTS, a framework that enables both accent control and conversion while preserving speaker identity. Specifically, we introduce an Accent Intensity Controller (AIC) that injects weighted language embeddings into the accent subspace, allowing smooth interpolation between accents and fine-grained modulation of accent strength at inference time. Experiments on the Indic Multilingual and L2-arctic datasets shows that CrossAccent-TTS achieves precise control of accent intensity, outperforming strong baselines in accent similarity and controllability by maintaining speaker similarity and naturalness.
\end{abstract}

\section{Introduction}
Speech accent refers to systematic variations in phonemes, rhythm, intonation, and linguistic structure, often providing cues about a speaker’s background, such as their geographic origin or native language \cite{moyer2013foreign}. However, disentangling accent from other speaker-specific attributes, such as pitch range, timbre, and vocal-tract characteristics, remains challenging \cite{zhou2026multi,liu2024controllable}, as accent is inherently embedded within individual speech patterns \cite{zhou2024accented}. Although accented TTS systems have numerous real-world applications, such as voice dubbing and conversational bots, they have not been a primary focus in mainstream TTS research. The ability to reproduce and control specific accents for a given speaker enables more authentic and personalized speech synthesis \cite{liu2024controllable}. In particular, in dubbing or voice acting, preserving a character’s L2 accent, i.e., the foreign accent that arises when speaking a non-native language, enhances realism and audience immersion. Conversely, overly strong or uncontrolled accents may reduce intelligibility, underscoring the need for fine-grained control over accent intensity to balance authenticity, clarity, and expressiveness.

While accent-controllable TTS remains underexplored, recent advances in LLM-based neural codec TTS have demonstrated scalable and flexible speech synthesis frameworks \cite{casanova2025low,casanova24_interspeech,sahipjohn24_interspeech}. These approaches combine the language modeling capabilities of LLMs with discrete acoustic representations derived from neural speech codecs. In this paradigm, speech is first encoded into a sequence of discrete acoustic tokens using a neural codec. A large language model is then trained to autoregressively generate these tokens conditioned on text, speaker identity, accent, or other controllable attributes. The generated token sequence is finally decoded by the decoder to reconstruct the final waveform.

\begin{figure*}[t]
  \centering
  \includegraphics[width=0.9\textwidth,height=0.6\textwidth,keepaspectratio]{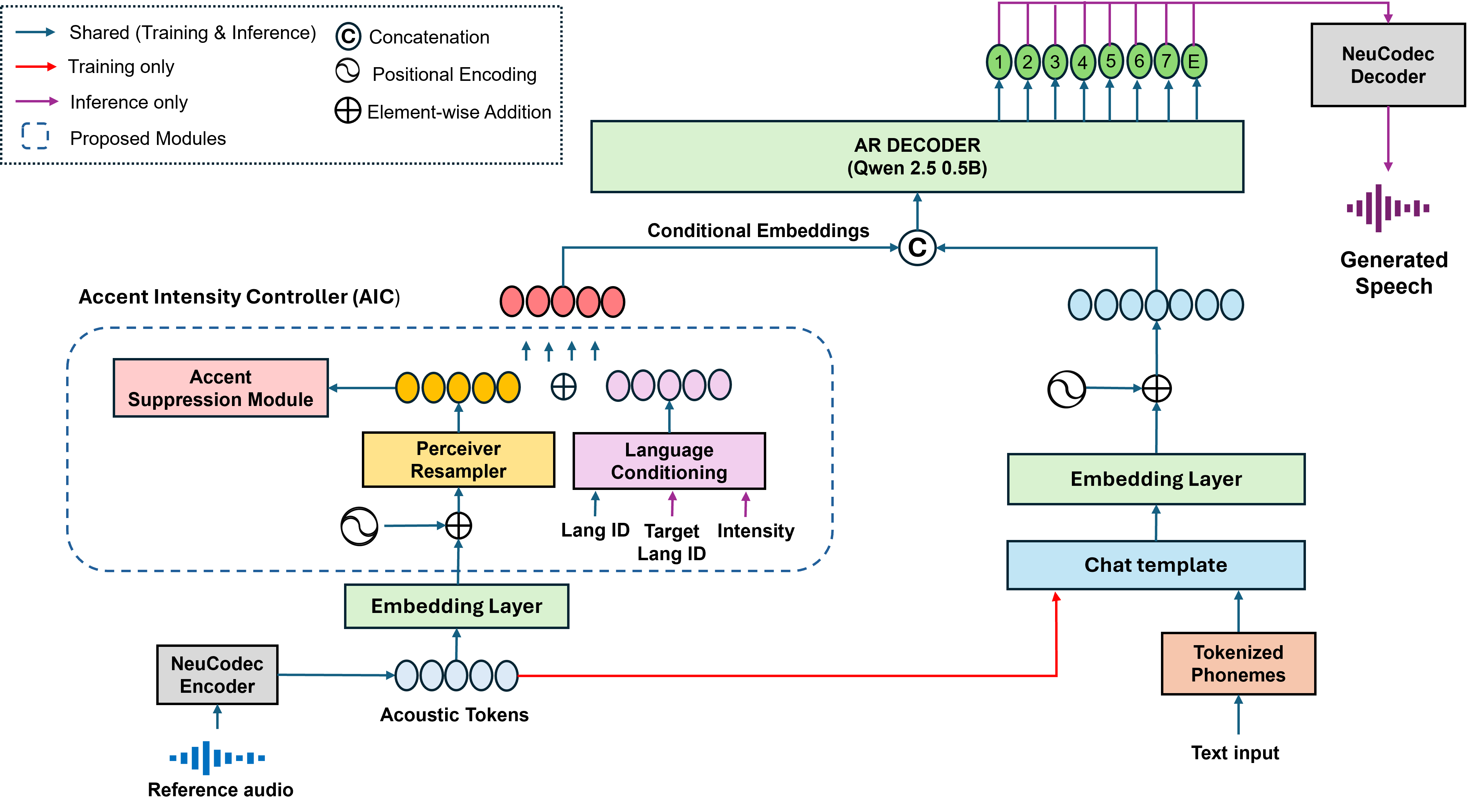}
  \caption{Model architecture of the proposed CrossAccent TTS}
  \label{fig:main_model}
\end{figure*}

To achieve accent-controllable TTS, we propose a Qwen-2.5-based model that uses a neural speech codec (Neucodec) to convert audio into discrete acoustic tokens. The model leverages the generative capability of LLMs to produce accent-conditioned acoustic representations while maintaining speaker identity. In contrast, we address the problem of controllable accent generation in text-to-speech, with particular emphasis on Indic languages, which remain low-resource and underexplored in this context. We additionally validate our method on foreign English accents to demonstrate generalization. Demo samples are available at URL\footnote{ Demo:~\url{https://research.sri-media-analysis.com/interspeech26-cross-accent-tts/}}. Our contributions are summarized as follows:

\begin{enumerate}
    \item We introduce an Accent Intensity Controller that enables continuous accent modulation at inference time without requiring accent-specific training data.
    \item We propose an Accent Suppression Module that promotes adversarial disentanglement of accent from speaker and style representations, thereby preserving speaker similarity.
    \item We evaluate our method on the Indic Multilingual and L2 ARCTIC datasets, demonstrating consistent accent controllability across Indian and foreign accents.
\end{enumerate}

The rest of the paper is organized as follows: Section \ref{sec:related} covers related work, followed by Section \ref{sec:method}, where we describe the proposed method. Section \ref{sec:experiments} details the training procedure and experiments, and finally, Section \ref{sec:conclusion} concludes the study.

\label{sec:related}
\section{Related Works}
Early Text-to-Speech (TTS) systems primarily focused on improving naturalness and speaker similarity, with limited attention given to accent modeling. Traditional approaches \cite{jia2018transfer} attempted to handle accent and speaker variation by conditioning multi-speaker TTS systems on speaker embeddings extracted using a speaker-verification encoder, which generates a fixed-dimensional representation from a short reference speech sample. These embeddings are then used to condition an acoustic model and a neural vocoder, enabling speech synthesis in the voices of both seen and unseen speakers. Since accent information is implicitly embedded within the learned speaker representations, such methods capture accent characteristics indirectly but do not provide explicit control over accent during inference. Other than this, Codec based TTS \cite{chen2024vall,ICLR2025_74a31a3b,lu2025robust} initially focused on extracting semantic tokens. 
More recently, single-stream acoustic tokenization approaches have been introduced \cite{wang2025spark,mousavi2025discrete}, which directly encode both linguistic and acoustic information, simplifying downstream TTS models and improving synthesis quality.

In the context of accent-controllable TTS, \cite{badlani2023vani} explored Indic-language-accented TTS, reducing L2 accent leakage by disentangling accent and speaker representations. However, comparable studies on accent controllability for Indic languages remain limited. To establish strong baselines for Indian languages, we adopt IndicF5\cite{varadhan2025phir} and XTTS-v2 \cite{casanova24_interspeech}, which have demonstrated competitive performance on Hindi and other related Indic languages. For English, \cite{xinyuan2025scalable} demonstrated accent control using large-scale datasets and data augmentation strategies. However, such approaches rely heavily on abundant labeled data and are not directly applicable in low-resource Indic settings. To evaluate under controlled low-resource conditions, we use the L2-ARCTIC dataset \cite{zhao2018l2} and compare against CVAE \cite{melechovsky2024accented} and GST-based \cite{wang2018style} baselines. Closest to our inspiration, \cite{zhang2023speak} proposed expanding language embeddings across all acoustic tokens to suppress L2 accent leakage in cross-lingual synthesis. We build on this direction by extending token-level language conditioning to accent-agnostic representations derived from speech codecs.

\section{Proposed Methodology}
\label{sec:method}
This section describes the proposed Cross-Accent TTS framework. An overview of the architecture is illustrated in \ref{fig:main_model}. The system consists of four main components: (i) speech tokenization using Neucodec, (ii) a Perceiver Resampler for speaker and style encoding with a fixed-length bottleneck, (iii) adversarial suppression of accent and language features, and (iv) explicit language embedding conditioning for controllable accent rendering, followed by auto-regressive acoustic token generation and Neucodec-based waveform synthesis.

\subsection{Speech Tokenization with Neucodec}
Neucodec \cite{julian2025finite} is a finite scalar quantization (FSQ)–based neural acoustic speech codec that operates at 50 tokens per second with 16 bits per token. It takes a 16 kHz input and reconstructs speech at 24 kHz using an upsampling decoder. The FSQ encoding scheme also provides improved robustness to bit-level errors. Ground-truth speech is first converted into discrete acoustic tokens using the Neucodec encoder. This encoder maps raw audio waveforms into a sequence of discrete latent codes that preserve perceptual speech quality while providing a compact token-based representation suitable for sequence modeling. These acoustic tokens are used to extract both speaker and style from the reference speech and to train acoustic token generation.
During training, a random chunk of acoustic tokens is sampled from each reference utterance. This random chunking strategy reduces overfitting to utterance-level linguistic content and encourages the Perceiver Resampler to focus on speaker-related characteristics rather than specific phonetic sequences.

\subsection{Accent Intensity Controller (AIC)}
The Accent Intensity Controller learns accent-agnostic speaker features while modeling accent features independently. Both are integrated as conditional embeddings to support explicit control over accent intensity.

\subsubsection{Perceiver Resampler}
The sampled acoustic token embeddings are fed into a Perceiver Resampler to obtain fixed-length speaker and style embeddings. Perceiver maps variable-length token sequences into a predefined set of latent vectors using cross-attention followed by self-attention layers. Before entering the encoder, learnable positional encodings are added to the reference token embeddings to preserve temporal structure. The fixed-length bottleneck constrains the information capacity of the input acoustic embeddings, encouraging the model to retain speaker-specific characteristics such as timbre and vocal quality, forming compact speaker and style embeddings of shape $(B, N_s, d)$. Since these embeddings are used to condition the autoregressive decoder to predict acoustic tokens from text, the bottleneck encourages the model to focus on speaker-related features while discarding linguistic and phonetic variations.\cite{jaegle2021perceiver}

\subsubsection{Adversarial learning with Accent Suppression Module}
To further suppress residual accent and language information in the speaker and style embeddings, we employ adversarial training with a gradient reversal layer (GRL). An auxiliary classifier is trained to predict the language or accent label from the speaker and style embeddings. During backpropagation, GRL inverts the gradient signal flowing from the classifier to the Perceiver Resampler, encouraging the encoder to remove accent- or language-discriminative information from the speaker and style embeddings. Formally, the Perceiver Resampler is optimized to minimize the TTS reconstruction loss while maximizing the classification error of the auxiliary accent/language classifier. This adversarial objective drives the encoder to produce accent- and language-invariant speaker and style embeddings during inference.

\subsubsection{Accent Control using Explicit Language Conditioning }
To explicitly control accent and language, we introduce a learned language embedding for each supported language or accent. A simple learnable embedding table is used to obtain a language embedding of shape $(B, 1, d)$. This embedding is then expanded across all latent slots of the speaker and style representations and added to each speaker and style embedding, yielding a combined representation of shape $(B, N_s, d)$. This expansion and addition mechanism increases the relative influence of language conditioning and enables the model to explicitly control linguistic and accent-related factors independently of the speaker and style.

\subsection{Autoregressive Acoustic Token Generation}
The combined speaker–language representation is provided as conditional input to an autoregressive decoder based on the Qwen architecture. We use Qwen 2.5 (0.5B) \cite{ahmed2025qwen} with the same configuration as the original model. The decoder jointly conditions on text token embeddings and the combined speaker–language embeddings to predict sequences of acoustic tokens. This autoregressive formulation allows the model to capture long-range dependencies and produce coherent acoustic token sequences conditioned on both linguistic content and controllable accent information. The final training objective is defined as:
\begin{equation}
L_{\text{total}} = L_{\text{decoder}} + \lambda_{\text{GRL}} \, L_{\text{GRL}}
\label{eq:total_loss}
\end{equation}
where $L_{\text{decoder}}$ is the autoregressive acoustic token prediction loss, and $L_{\text{GRL}}$ is the adversarial classification loss used for accent and language suppression and $\lambda_{\text{GRL}}$ is a weighting factor that controls the relative contribution of the GRL loss in the total training objective.

\section{Training and Evaluation Setup}
\label{sec:experiments}
This section describes the experimental setup, evaluation metrics, and results to assess the proposed model's performance.
\subsection{Datasets}
We conduct experiments on two datasets, Indic Multilingual and L2 ARCTIC, to evaluate performance in low-resource and second-language (L2) accent scenarios.

\subsubsection{Indic Multilingual Dataset}
We use a multilingual speech corpus comprising Hindi, Telugu, Tamil, Bengali, Marathi, and English. The dataset contains 636 hours of internal in-house Indic speech data and 350 hours of custom-split speech from the Emilia Yodas dataset \cite{he2025emilia}, totaling approximately 986 hours of speech. All audios are resampled to 16 kHz for Neucodec encoding. Language labels are used to explicitly condition language embeddings and to perform adversarial training.

\subsubsection{L2 Arctic Dataset}
 The L2 Arctic corpus comprises 27 hours of non-native English speech from 24 speakers, with six distinct accent backgrounds. The dataset provides aligned text and speech with annotated speaker and native-language labels, making it suitable for evaluating accent neutrality and for controllable accent rendering in L2 English synthesis.

\subsection{Training Setup}
The proposed model is trained end-to-end using paired text and speech data. 
The text is converted to International Phonetic Alphabets (IPA) and tokenized using a shared tokenizer that maps speech tokens. The Perceiver Resampler uses $N_s = 32$ latent slots and embedding dimension $d = 768$. 
The autoregressive decoder is initialized from Qwen 2.5 (0.5B) and fine-tuned for acoustic token generation. The auxiliary accent/language classifier is trained jointly with the main model using the accent suppression module. The total training loss is defined as shown in Eq. \ref{eq:total_loss}, where $\lambda_{\text{GRL}}$ controls the strength of adversarial accent and language suppression. 

We set $\lambda_{\text{GRL}} = 0.1$ in all experiments. The decoder is first trained on the full 986-hour multilingual dataset for 5 epochs, then fine-tuned on the L2 Arctic dataset with a reduced learning rate for 3 additional epochs. During inference, the reference speech is used to extract speaker and style information, and multiple language embeddings can be linearly combined to control accent intensity. Specifically, given language embeddings $e_{\text{lang}_1}$ and $e_{\text{lang}_2}$, a weighted combination is used,
\begin{equation}
\lambda e_{\text{lang}_1} + (1-\lambda) e_{\text{lang}_2}
\end{equation}
where $\lambda \in [0,1]$ controls the relative contribution. The predicted acoustic tokens are passed through the Neucodec decoder to reconstruct the time-domain speech waveform.

\subsection{Evaluation Metrics}
We evaluate the model's performance using both objective and subjective metrics.

\subsubsection{Objective Metrics}
To obtain accent embeddings for evaluation, we use the GenAID model originally introduced in the Accent Box \cite{zhao2018l2} for L2 English accent analysis. For Indic languages, we fine-tune the same model using approximately 100k samples from 11 Indian languages and around 3,300 speakers from the Indic Voices dataset \cite{javed2024indicvoices} for two epochs. Accent embeddings extracted from this fine-tuned model are used for Indic accent evaluations. We report the following objective measures:
\begin{itemize}
    \item \textbf{Accent Similarity:} Measures how closely synthesized speech matches the target accent. 
    We compute cosine similarity between the accent embeddings of generated speech and those of ground-truth speech in the target accent. Higher values indicate better accent rendering.
    
    \item \textbf{Accent Leakage:} Measures how much accent information from the reference audio is preserved when synthesizing in a different target accent. We compute the cosine similarity between the accent embeddings of the reference and generated audio. 
    Lower values indicate better suppression of source accent.    
    
    \item \textbf{UTMOS \cite{saeki2022utmos}:} This metric is used to assess overall speech quality.
    
    \item \textbf{Speaker Similarity:} Computed using a pretrained speaker verification model (Resemblyzer) \cite{louppe2019resemblyzer} to assess speaker identity preservation.
\end{itemize}

\subsubsection{Subjective Metrics}
Subjective listening tests are conducted using the Mean Opinion Score (MOS) to evaluate how closely the synthesized speech matches the target accent. Listening tests are conducted with 20 participants (aged between 23 to 35 with no known hearing impairments).

\subsection{Evaluation Results}
\subsubsection{Objective Results}
Table \ref{tab:accent_results} and Table \ref{tab:l2_arctic_results} report the objective evaluation results on the Indic multilingual and L2 Arctic datasets. The proposed method achieves reduced accent leakage compared to baselines, demonstrating more effective suppression of source accent in accent-neutral synthesis. When explicit language embeddings are applied, accent similarity increases in a controlled manner, confirming the effectiveness of the proposed accent control mechanism.
The proposed method also maintains speaker similarity scores comparable to the external speaker embedding baseline, indicating that speaker identity is preserved despite adversarial accent suppression. The UTMOS results further show that overall speech quality is maintained or improved relative to baseline systems.
\begin{table}[h]
    \setlength{\abovecaptionskip}{2pt} 
  \setlength{\belowcaptionskip}{1pt} 
  \setlength{\textfloatsep}{1pt} 
  \centering
  \small
  \setlength{\tabcolsep}{4pt}
  
\caption{Objective evaluation results on Indian accent conversion ($\uparrow$ higher is better, $\downarrow$ lower is better).}
  \begin{tabular}{l c c c c}
    \toprule
    \textbf{Model} & \textbf{UTMOS $\uparrow$} & \textbf{AccLeak $\downarrow$} & \textbf{AccSim $\uparrow$} & \textbf{SpkSim $\uparrow$} \\
    \midrule
    IndicF5   & 2.817 & 0.312 & 0.312 & \textbf{0.843} \\
    XTTS\_v2  & 3.168 & 0.284 & 0.284 & 0.832 \\
    Proposed  & \textbf{3.181} & \textbf{0.203} & \textbf{0.371} & 0.842 \\
    \bottomrule
  \end{tabular}

  \label{tab:accent_results}
\end{table}
\begin{table}[h]
    \setlength{\abovecaptionskip}{1pt} 
  \setlength{\belowcaptionskip}{1pt} 
  \setlength{\textfloatsep}{1pt} 
  \centering
  \small
  \setlength{\tabcolsep}{4pt}
\caption{Objective evaluation results on L2 Arctic dataset with English accents ($\uparrow$ higher is better, $\downarrow$ lower is better).}
  \begin{tabular}{l c c c c}

    \toprule
    \textbf{Model} & \textbf{UTMOS $\uparrow$} & \textbf{AccLeak $\downarrow$} & \textbf{AccSim $\uparrow$} & \textbf{SpkSim $\uparrow$} \\
    \midrule
    CVAE-L        & 2.810 & 0.487 & 0.612 & 0.677 \\
    CVAE-NL       & 2.714 & 0.530 & 0.491 & 0.673 \\
    GST           & 3.044 & 0.544 & 0.670 & \textbf{0.732} \\
    Proposed      & \textbf{4.001} & \textbf{0.439} & \textbf{0.686} & 0.693 \\
    \bottomrule
  \end{tabular}

  \label{tab:l2_arctic_results}
\end{table}
\subsubsection{Subjective Results}
\begin{figure}[h]
  \centering
  \begin{minipage}{0.50\columnwidth}
    \centering
    \includegraphics[width=\linewidth]{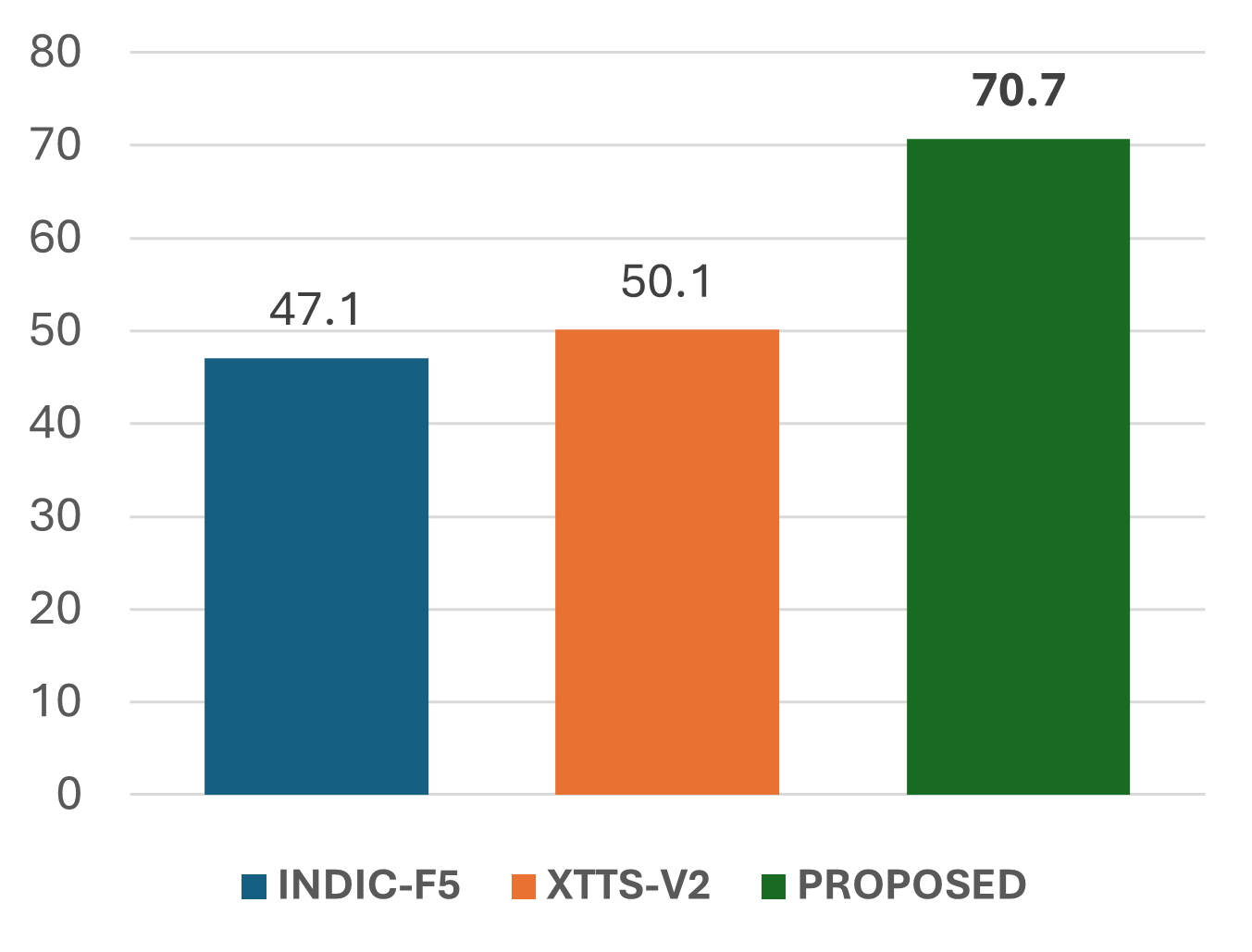}
    \caption{Accent Similarity MOS of Indic Accents}
    \label{fig:IndicAccents}
  \end{minipage}
  \hfill
  \begin{minipage}{0.48\columnwidth}
    \centering
    \includegraphics[width=\linewidth]{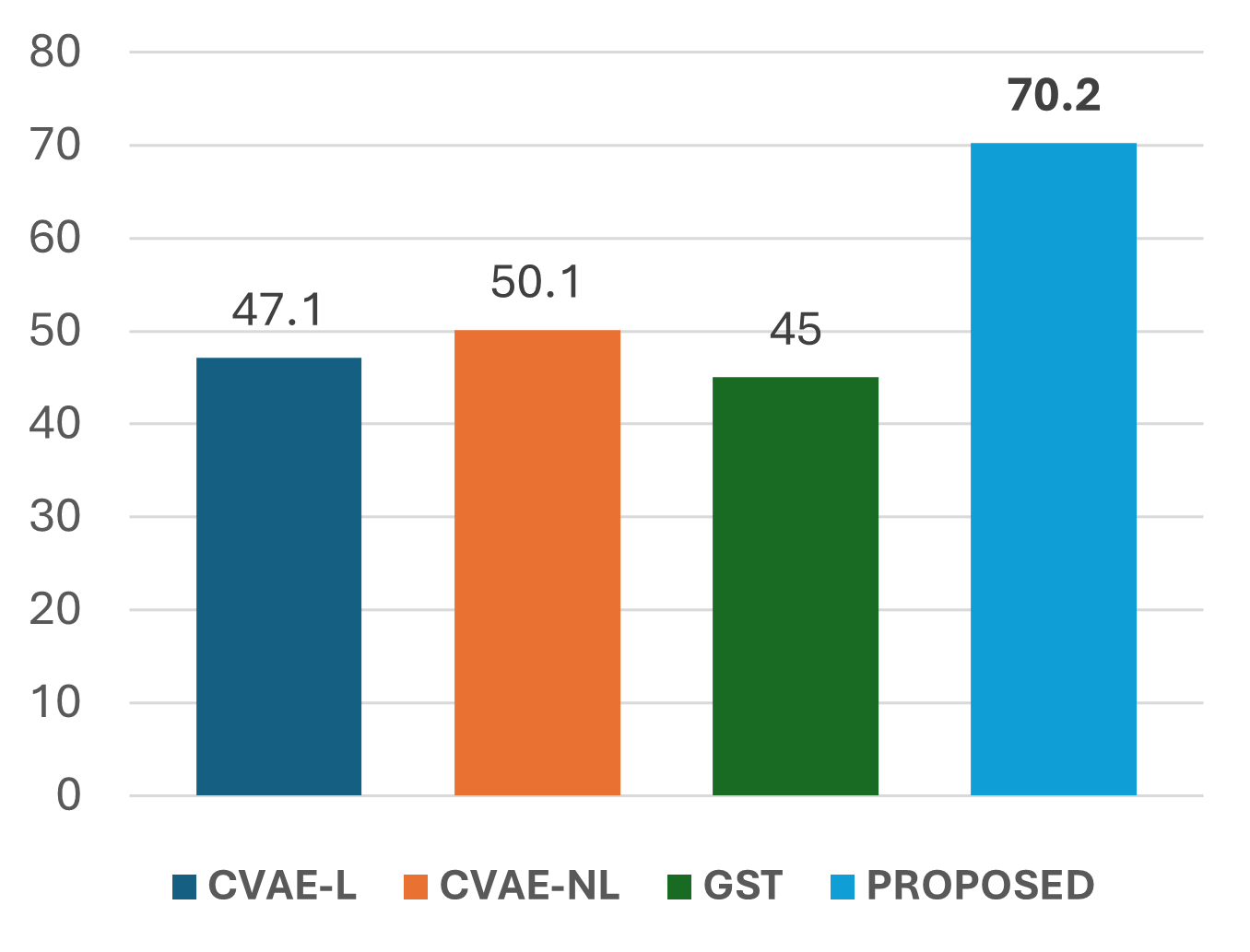}
    \caption{Accent Similarity MOS of Foreign Accents}
    \label{fig:ForeignAccents}
  \end{minipage}
\end{figure}

Figure \ref{fig:IndicAccents} and Figure \ref{fig:ForeignAccents} represent the subjective listening test results for Indian accents and foreign accents, respectively. The proposed method achieves higher MOS scores for accent similarity compared to baseline systems. In accent-controlled settings, listeners rate the synthesized speech as more similar to the target accent, confirming that weighted language embedding enables perceptually meaningful accent control.




\subsection{Accent-intensity Control Analysis}
\begin{figure}[h!]
  \centering
  \includegraphics[width=0.75\columnwidth]{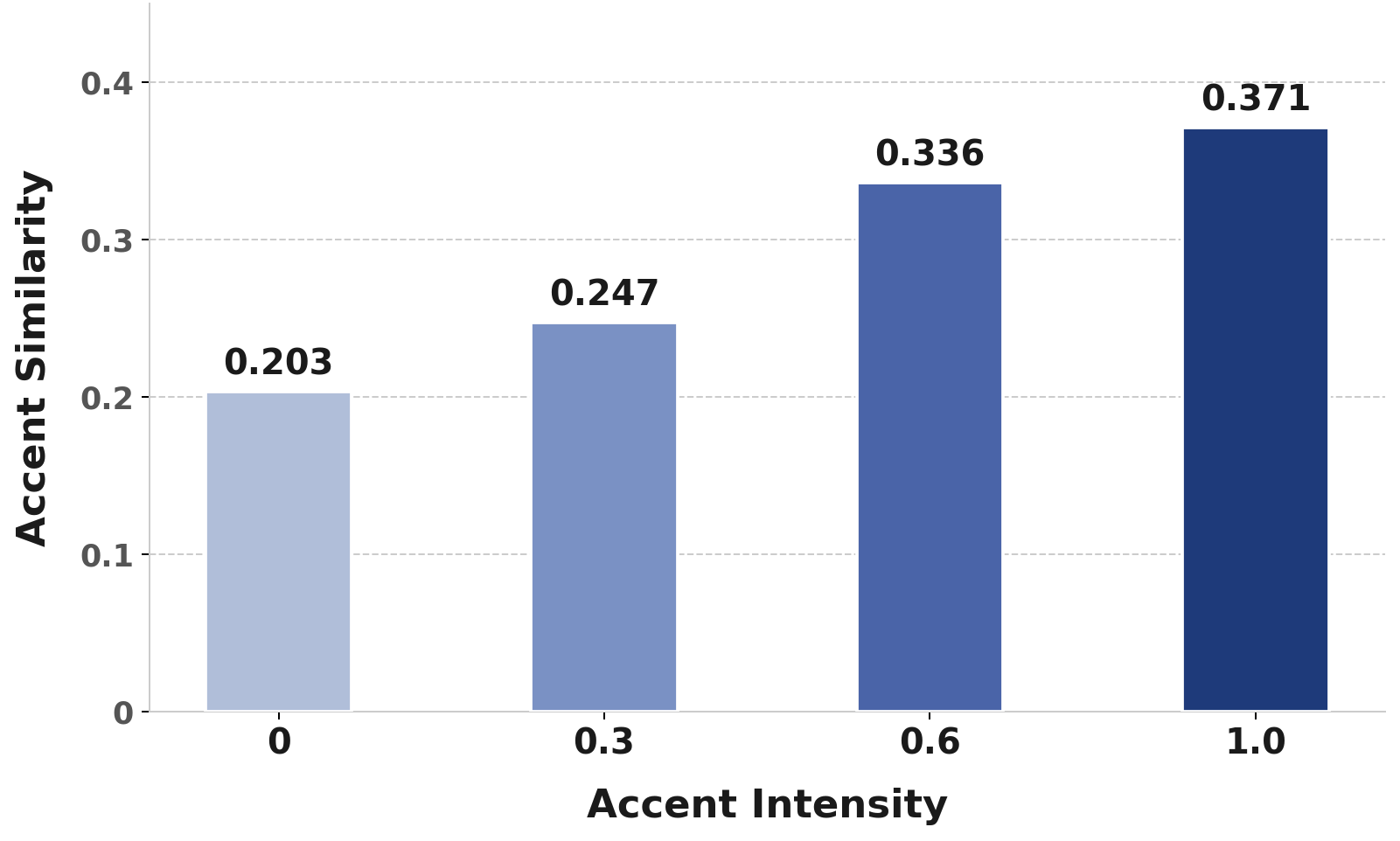}
  \caption{Effect of Accent Intensity on Accent Similarity score}
  \label{fig:Effect of Accent Intensity on Cosine Accent Similarity score}
\end{figure}

To assess accent-intensity control, we synthesize samples using cross-lingual reference audio, scaling the accent intensity to 0, 0.3, 0.6, and 1.0. Accent embeddings from the synthesized speech are extracted using GenAID, and accent similarity scores are computed with the mentioned intensity scores. From the Figure \ref{fig:Effect of Accent Intensity on Cosine Accent Similarity score}, it is evident that as accent intensity increases, the accent similarity score also rises, demonstrating effective and controllable accent modulation.

\section{Conclusion}
\label{sec:conclusion}
This paper presents CrossAccent TTS, a cross-lingual, accent-controllable text-to-speech framework that enables both accent conversion and continuous modulation of accent intensity while preserving speaker identity. The proposed Accent Intensity Controller injects weighted language embeddings into disentangled representations, allowing smooth interpolation across accents and dynamic control during inference. Extensive objective and subjective evaluations demonstrate that the proposed approach consistently outperforms strong baselines in accent similarity and controllability, while maintaining speaker similarity and naturalness. The effectiveness of the learned accent representations is further supported by accent-intensity control analysis. Overall, the proposed framework provides a practical and scalable solution for accent-controllable speech synthesis in low-resource and multilingual settings.


\section{Generative AI Use Disclosure}
Generative AI tools were used solely for language refinement, grammar correction, and improving the overall clarity and readability of the manuscript. These tools assisted in polishing and structuring the text but were not used to generate or design the core research ideas, methodology, experimental setup, analysis, or conclusions presented in this work. All scientific contributions, technical implementations, and interpretations were developed and validated by the authors. In accordance with policy guidelines, no generative AI system is listed as a co-author, and all authors take full responsibility and accountability for the content and integrity of this paper.

\bibliographystyle{IEEEtran}
\bibliography{mybib}

@article{jia2018transfer,
  title={Transfer learning from speaker verification to multispeaker text-to-speech synthesis},
  author={Jia, Ye and Zhang, Yu and Weiss, Ron and Wang, Quan and Shen, Jonathan and Ren, Fei and Nguyen, Patrick and Pang, Ruoming and Lopez Moreno, Ignacio and Wu, Yonghui and others},
  journal={Advances in neural information processing systems},
  volume={31},
  year={2018}
}

@article{chen2024vall,
  title={Vall-e 2: Neural codec language models are human parity zero-shot text to speech synthesizers},
  author={Chen, Sanyuan and Liu, Shujie and Zhou, Long and Liu, Yanqing and Tan, Xu and Li, Jinyu and Zhao, Sheng and Qian, Yao and Wei, Furu},
  journal={arXiv preprint arXiv:2406.05370},
  year={2024}
}

@book{moyer2013foreign,
  title={Foreign accent: The phenomenon of non-native speech},
  author={Moyer, Alene},
  year={2013},
  publisher={Cambridge University Press}
}

@article{zhou2026multi,
  title={Multi-scale accent modeling and disentangling for multi-speaker multi-accent text-to-speech synthesis},
  author={Zhou, Xuehao and Zhang, Mingyang and Zhou, Yi and Wu, Zhizheng and Li, Haizhou},
  journal={IEEE Transactions on Audio, Speech and Language Processing},
  year={2026},
  publisher={IEEE}
}

@article{zhou2024accented,
  title={Accented text-to-speech synthesis with limited data},
  author={Zhou, Xuehao and Zhang, Mingyang and Zhou, Yi and Wu, Zhizheng and Li, Haizhou},
  journal={IEEE/ACM Transactions on Audio, Speech, and Language Processing},
  volume={32},
  pages={1699--1711},
  year={2024},
  publisher={IEEE}
}

@inproceedings{ICLR2025_74a31a3b,
 author = {Wang, Yuancheng and Zhan, Haoyue and Liu, Liwei and Zeng, Ruihong and Guo, Haotian and Zheng, Jiachen and Zhang, Qiang and Zhang, Xueyao and Zhang, Shunsi and Wu, Zhizheng},
 booktitle = {International Conference on Learning Representations (ICLR)},
 pages = {47127--47150},
 title = {MaskGCT: Zero-Shot Text-to-Speech with Masked Generative Codec Transformer},
 url = {https://proceedings.iclr.cc/paper_files/paper/2025/file/74a31a3b862eb7f01defbbed8e5f0c69-Paper-Conference.pdf},
 year = {2025}
}

@inproceedings{casanova2025low,
  title={Low frame-rate speech codec: a codec designed for fast high-quality speech llm training and inference},
  author={Casanova, Edresson and Langman, Ryan and Neekhara, Paarth and Hussain, Shehzeen and Li, Jason and Ghosh, Subhankar and Juki{\'c}, Ante and Lee, Sang-gil},
  booktitle={ICASSP 2025-2025 IEEE International Conference on Acoustics, Speech and Signal Processing (ICASSP)},
  pages={1--5},
  year={2025},
  organization={IEEE}
}

@inproceedings{lu2025robust,
  title={Robust neural codec language modeling with phoneme position prediction for zero-shot tts},
  author={Lu, Chunhui and Wen, Xue and Song, Liming and Oh, Junkwang},
  booktitle={Proc. Interspeech 2025},
  pages={2475--2479},
  year={2025}
}

@article{mousavi2025discrete,
  title={Discrete audio tokens: More than a survey!},
  author={Mousavi, Pooneh and Maimon, Gallil and Moumen, Adel and Petermann, Darius and Shi, Jiatong and Wu, Haibin and Yang, Haici and Kuznetsova, Anastasia and Ploujnikov, Artem and Marxer, Ricard and others},
  journal={arXiv preprint arXiv:2506.10274},
  year={2025}
}

@article{wang2025spark,
  title={Spark-tts: An efficient llm-based text-to-speech model with single-stream decoupled speech tokens},
  author={Wang, Xinsheng and Jiang, Mingqi and Ma, Ziyang and Zhang, Ziyu and Liu, Songxiang and Li, Linqin and Liang, Zheng and Zheng, Qixi and Wang, Rui and Feng, Xiaoqin and others},
  journal={arXiv preprint arXiv:2503.01710},
  year={2025}
}

@inproceedings{badlani2023vani,
  title={VANI: Very-lightweight accent-controllable TTS for native and non-native speakers with identity preservation},
  author={Badlani, Rohan and Arora, Akshit and Ghosh, Subhankar and Valle, Rafael and Shih, Kevin J and Santos, Jo{\~a}o Felipe and Ginsburg, Boris and Catanzaro, Bryan},
  booktitle={ICASSP 2023-2023 IEEE International Conference on Acoustics, Speech and Signal Processing (ICASSP)},
  pages={1--2},
  year={2023},
  organization={IEEE}
}

@inproceedings{casanova24_interspeech,
  title     = {{XTTS: a Massively Multilingual Zero-Shot Text-to-Speech Model}},
  author    = {Edresson Casanova and Kelly Davis and Eren Gölge and Görkem Göknar and Iulian Gulea and Logan Hart and Aya Aljafari and Joshua Meyer and Reuben Morais and Samuel Olayemi and Julian Weber},
  year      = {2024},
  booktitle = {{Interspeech 2024}},
  pages     = {4978--4982},
  doi       = {10.21437/Interspeech.2024-2016},
  issn      = {2958-1796},
}

@article{xinyuan2025scalable,
  title={Scalable Controllable Accented TTS},
  author={Xinyuan, Henry Li and Cai, Zexin and Garg, Ashi and Duh, Kevin and Garc{\'\i}a-Perera, Leibny Paola and Khudanpur, Sanjeev and Andrews, Nicholas and Wiesner, Matthew},
  journal={arXiv preprint arXiv:2508.07426},
  year={2025}
}

@inproceedings{zhao2018l2,
  title     = {{L2-ARCTIC: A Non-native English Speech Corpus}},
  author    = {Guanlong Zhao and Sinem Sonsaat and Alif Silpachai and Ivana Lucic and Evgeny Chukharev-Hudilainen and John Levis and Ricardo Gutierrez-Osuna},
  year      = {2018},
  booktitle = {{Interspeech 2018}},
  pages     = {2783--2787},
  doi       = {10.21437/Interspeech.2018-1110},
  issn      = {2958-1796},
}

@inproceedings{melechovsky2024accented,
  title={Accented text-to-speech synthesis with a conditional variational autoencoder},
  author={Melechovsky, Jan and Mehrish, Ambuj and Sisman, Berrak and Herremans, Dorien},
  booktitle={TENCON 2024-2024 IEEE Region 10 Conference (TENCON)},
  pages={343--346},
  year={2024},
  organization={IEEE}
}

@article{liu2024controllable,
  title={Controllable accented text-to-speech synthesis with fine and coarse-grained intensity rendering},
  author={Liu, Rui and Sisman, Berrak and Gao, Guanglai and Li, Haizhou},
  journal={IEEE/ACM Transactions on Audio, Speech, and Language Processing},
  volume={32},
  pages={2188--2201},
  year={2024},
  publisher={IEEE}
}

@inproceedings{wang2018style,
  title={Style tokens: Unsupervised style modeling, control and transfer in end-to-end speech synthesis},
  author={Wang, Yuxuan and Stanton, Daisy and Zhang, Yu and Ryan, RJ-Skerry and Battenberg, Eric and Shor, Joel and Xiao, Ying and Jia, Ye and Ren, Fei and Saurous, Rif A},
  booktitle={International conference on machine learning},
  pages={5180--5189},
  year={2018},
  organization={PMLR}
}

@article{zhang2023speak,
  title={Speak foreign languages with your own voice: Cross-lingual neural codec language modeling},
  author={Zhang, Ziqiang and Zhou, Long and Wang, Chengyi and Chen, Sanyuan and Wu, Yu and Liu, Shujie and Chen, Zhuo and Liu, Yanqing and Wang, Huaming and Li, Jinyu and others},
  journal={arXiv preprint arXiv:2303.03926},
  year={2023}
}

@article{julian2025finite,
  title={Finite Scalar Quantization Enables Redundant and Transmission-Robust Neural Audio Compression at Low Bit-rates},
  author={Julian, Harry and Beeson, Rachel and Konathala, Lohith and Ulin, Johanna and Gao, Jiameng},
  journal={arXiv preprint arXiv:2509.09550},
  year={2025}
}

@inproceedings{jaegle2021perceiver,
  title={Perceiver: General perception with iterative attention},
  author={Jaegle, Andrew and Gimeno, Felix and Brock, Andy and Vinyals, Oriol and Zisserman, Andrew and Carreira, Joao},
  booktitle={International conference on machine learning},
  pages={4651--4664},
  year={2021},
  organization={PMLR}
}

@article{ahmed2025qwen,
  title={Qwen 2.5: A comprehensive review of the leading resource-efficient llm with potentioal to surpass all competitors},
  author={Ahmed, Imtiaz and Islam, Sadman and Datta, Partha Protim and Kabir, Imran and Chowdhury, Naseef Ur Rahman and Haque, Ahshanul},
  journal={Authorea Preprints},
  year={2025},
  publisher={Authorea}
}

@inproceedings{javed2024indicvoices,
  title={Indicvoices: Towards building an inclusive multilingual speech dataset for indian languages},
  author={Javed, Tahir and Nawale, Janki and George, Eldho and Joshi, Sakshi and Bhogale, Kaushal and Mehendale, Deovrat and Sethi, Ishvinder and Ananthanarayanan, Aparna and Faquih, Hafsah and Palit, Pratiti and others},
  booktitle={Findings of the Association for Computational Linguistics: ACL 2024},
  pages={10740--10782},
  year={2024}
}

@misc{louppe2019resemblyzer,
  author       = {Louppe, Gilles},
  title        = {Resemblyzer: Voice Encoder for Speaker Verification},
  year         = {2019},
  howpublished = {\url{https://github.com/resemble-ai/Resemblyzer}},
  note         = {Accessed: 2026-02-19}
}

@article{saeki2022utmos,
  title={Utmos: Utokyo-sarulab system for voicemos challenge 2022},
  author={Saeki, Takaaki and Xin, Detai and Nakata, Wataru and Koriyama, Tomoki and Takamichi, Shinnosuke and Saruwatari, Hiroshi},
  journal={arXiv preprint arXiv:2204.02152},
  year={2022}
}

@article{varadhan2025phir,
  title={Phir Hera Fairy: An English Fairytaler is a Strong Faker of Fluent Speech in Low-Resource Indian Languages},
  author={Varadhan, Praveen Srinivasa and Anand, Srija and Siddhartha, Soma and Khapra, Mitesh M},
  journal={arXiv preprint arXiv:2505.20693},
  year={2025}
}

@article{he2025emilia,
  title={Emilia: A large-scale, extensive, multilingual, and diverse dataset for speech generation},
  author={He, Haorui and Shang, Zengqiang and Wang, Chaoren and Li, Xuyuan and Gu, Yicheng and Hua, Hua and Liu, Liwei and Yang, Chen and Li, Jiaqi and Shi, Peiyang and others},
  journal={IEEE Transactions on Audio, Speech and Language Processing},
  year={2025},
  publisher={IEEE}
}

@inproceedings{sahipjohn24_interspeech,
  title     = {{DubWise: Video-Guided Speech Duration Control in Multimodal LLM-based Text-to-Speech for Dubbing}},
  author    = {Neha Sahipjohn and Ashishkumar Gudmalwar and Nirmesh Shah and Pankaj Wasnik and Rajiv Ratn Shah},
  year      = {2024},
  booktitle = {{Interspeech 2024}},
  pages     = {2960--2964},
  doi       = {10.21437/Interspeech.2024-1700},
  issn      = {2958-1796},
}

\end{document}